\def\year{2016}\relax
\documentclass[letterpaper]{article}
\usepackage{aaai16}
\usepackage{times}
\usepackage{helvet}
\usepackage{courier}
\usepackage{balance}
\begin{document}
%
\title{Creating Interactive Behaviors in Early Sketch by\\ Recording and Remixing Crowd Demonstrations}


\author{
  \parbox[t]{14cm}{\centering
  Sang Won Lee$^1$, Yiwei Yang$^1$, Shiyan Yan$^2$, Yujin Zhang$^1$, Isabelle Wong$^1$, Zhengxi Tan$^1$, Miles McGruder$^1$, Christopher Homan$^3$, Walter S. Lasecki$^{1,2}$\\
  }
  \vspace{0.5pc}\\
  \parbox[t]{5.5cm}{\centering
    Computer Science \& Engineering$^1$\\
    University of Michigan, Ann Arbor\\
    \{snaglee,wlasecki\}@umich.edu\\
  }
  \parbox[t]{5.5cm}{\centering
    School of Information$^2$\\
    University of Michigan, Ann Arbor\\
    shiyansi@umich.edu\\
  }
  \parbox[t]{6cm}{\centering
    Department of Computer Science$^3$\\
    Rochester Institute of Technology\\
    cmh@cs.rit.edu\\
  }
}
\vspace{-20pt}

\maketitle

In the early stages of designing graphical user interfaces (GUIs), the look (appearance) can be easily presented by sketching, but the feel (interactive behaviors) cannot, and often requires an accompanying description of how it works~\cite{myers2008designers}.
We propose to use crowdsourcing to augment early sketches with interactive behaviors generated, used, and reused by collective ``wizards-of-oz'' as opposed to a single wizard as in prior work~\cite{Davis:2007:SWO:1294211.1294233}. 
This demo presents an extension of Apparition~\cite{Lasecki:2015:ACU:2702123.2702565}, a crowd-powered prototyping tool that allows end users to create functional GUIs using speech and sketch.
In Apparition, crowd workers collaborate in real-time on a shared canvas to refine the user-requested sketch interactively, and with the assistance of the end users.
Our demo extends this functionality to let crowd workers ``demonstrate'' the canvas changes that are needed for a behavior and refine their demonstrations to improve the fidelity of interactive behaviors.
The system then lets workers ``remix'' these behaviors to make creating future behaviors more efficient.

\vspace{-5pt}
\section{Challenges and Contributions}

Demonstrating interactive behaviors in early prototypes is challenging for multiple reasons. 
Interactive behaviors involve the dynamic transformation of multiple user interface elements, meaning they cannot be easily described by one static image. 
Furthermore, to accurately describe the behaviors often requires one to resort to abstract language: e.g., logic, states, and constraints. 
While there exist a number of commercial products (e.g., Adobe FireWorks, Sketch, InVision) to create interactive mock-ups, using such tools during the early iterations of a design project is time-consuming compared to sketching on paper. 
Additionally, commercial production often provides only a fixed set of behaviors that are tuned to support specific types of interactions (e.g., website wire-framing) and is often too limited for complex behaviors. 
For example, animating the behaviors of a game character (e.g. Super Mario) or how the enemy characters (e.g. turtles in Super Mario) respond to the other element's changes (e.g., Mario bouncing on them) cannot be done easily with such products. 
A more traditional approach to demonstrating such behaviors---one that is not as limited to specific types of interactions---is to use paper cut-outs. 
But this is limited by the effort needed to produce the cutouts. Moreover, the resulting demonstrations of system behavior cannot be as easily saved for future use.

In this work, we leverage crowd workers to demonstrate the interactive behaviors in real time (tens of seconds), and provide system support to document and retain the interactive behaviors for further iterations, communication, and development. 
We adopt record-and-replay features from existing tools for animations \cite{davis2008k}, interactive sketch \cite{obrenovic2011sketching}  and interactive illustration composition \cite{kazi2014kitty}, but our system eliminates the end users' needs to learn new sketching software. Instead we let them create interactive sketches by verbally describing what should happen and the context of the system is focused on low-mid fidelity sketching tools for the agile design process, rather than creating nice looking interactive animations. 
Our approach is challenging due to i) varying levels of crowd expertise on such tools; ii) the temporary nature of crowd workers, who may freely come and go throughout a session; and iii) the nature of real-time groupware, which depends on coordination and awareness~\cite{gutwin2002descriptive}. 

\vspace{0.3pc}
\noindent We showcase the following in this demonstration:

\vspace{-0.2pc}
\begin{itemize}
\item The idea of ``demonstrate-and-remix'' in early-stage sketch prototyping for creating interactive behaviors
\vspace{-0.1pc}
\item Methods for maintaining knowledge about behaviors
\vspace{-0.1pc}
\item Methods and tools for coordinating crowds in real-time
\end{itemize}

\vspace{-5pt}
\section{Design and Implementation}

The complexity of interactive behaviors can vary highly: from Super Mario bouncing on a Turtle to marking an item as done in a ``TODO'' list.
Typically, interactive behaviors have three properties: $P1$: what \emph{triggers} the interactive behavior, $P2$: the visual state changes made to the GUI by the interactive behaviors, and  $P3$: non-visible states and relationships that may affect current and future behaviors.

For instance, in a game setting when Super Mario bounces on a turtle, the end-user's verbal description could be: ``When Super Mario jumps and lands on top of an enemy turtle [$P1$], the turtle should flip upside down, and hide its head and legs in the shell [$P2$]; the shell can then be used as a weapon once it lands on the ground [$P3$]''. 
The interactive behavior here includes interaction (button press) state recognition (triggering condition: Mario bouncing on the turtle), animated behavior (turtle flipping and hiding), and abstract relationships (gravity, behavior conditions). 
Variation makes automating and composing such interactive behaviors difficult without programming (where the time and effort required is inappropriate for early prototyping). 
However,  it is relatively easy for a crowd worker to understand the triggering condition ($P1$), demonstrate what happens by changing objects on canvas ($P2$) and the key abstract relationships ($P3$). 
Existing approaches require text descriptions to explain $P1$--$P3$, but complex or synchronous visual changes ($P2$) are difficult to capture in linear text alone.
We use a ``demonstrate-and-remix'' approach for \emph{showing} visual changes ($P2$) and  maintain a collective memory ($P1$, $P3$).

\balance

\vspace{-1pt}
\subsection{Demonstrate-and-Remix} 
\vspace{-1pt}

Constructing visual state changes of interactive behaviors in Apparition involves two (repeatable) phases: \emph{demonstrate} and \emph{remix}. 
An interactive behavior can be demonstrated as a series of element-wise transformations on the shared canvas. 
%
To create an interactive behavior, a crowd worker demonstrates (or performs) what is supposed to happen. 
In our Mario-turtle example, a crowd worker can a) rotate the turtle to make it upside down, b) make the turtle bounce as part of a dying gesture by drag-and-drop, c) import a new image of a turtle with its legs and head hidden, d) and delete the original image of the turtle. 
While the crowd worker demonstrates what needs to happen, Apparition records all interaction the crowd workers perform on the canvas and organizes them into sequences of ``operations'' on each element (here: rotation, translation, creation and deletion, respectively). 

In the ``remix'' phase, these demonstrations are available for crowd workers to edit and create new interactive behaviors. 
This process is analogous to a DJ chopping, editing, processing and arranging audio samples to make music. 
The idea of remix in real-time collaboration draws upon our previous works in collaborative improvisation, where musicians algorithmically remix short musical patterns \cite{lee2012evaluating} and musical notation \cite{lee2013real}, or programmers share a program state \cite{lee2014communication}.
In our system, each operation is available as a set of visual blocks in a timeline editor. 
The length of each block represents the time it takes to replay the recorded operation.
The system provides a set of remix functions for a crowd worker to edit. 
For example, \textit{compress/stretch} operation can change the duration of an operation (or even make each operation instantaneous), \textit{trim/skip} can cut some changes that are not desirable within an operation,
\textit{normalize/smooth/ease-in-out} can refine the temporal execution of the demonstration, and \textit{reverse} can create a backwards version of the demonstration. 
The remix functions includes visually editing the trajectory like \textit{resize},\textit{rotate} and generative functions like \textit{clone}, \textit{apply}. 

In the final step, the (remixed) operations can be placed in the timeline so that a worker can temporally control execution time, intervals, overlaps and the order of operations, as one behavior. 
In our Mario-turtle example, a crowd worker can decide to make the operation (a), (c), (d) instant and compress b) to happen in 1 sec and place the blocks in the order: (a)-(b)-(d)-(c).
Or if the worker wants to immediately hide its legs and head, one can put (d) and (c) in front and \textit{apply} a) and b) to the imported image instead of the original turtle image.
This arrangement on the timeline editor concatenates the operations into one single behavior that can be reproduced.
The demonstrate-and-remix approach maintains a simple and easy process of composing an interactive behavior so that crowd workers can quickly create complex behaviors.

\vspace{-3pt}
\subsection{Retaining the Interactive Behaviors} 
\vspace{-2pt}
Once the visual changes of one behavior have been reproduced, a crowd worker can observe the sketch and find the trigger for the specified behaviors to make the sketch interactive (e.g., observing when Super Mario makes contact with a turtle). 
In this scenario, P1: (triggering conditions) and P3: (abstract relationships) are left to crowd workers' judgements, but the problem ($P1$,$P3$) is that this cannot be easily automated, meaning the system cannot retain the behaviors unless the same crowd worker(s) are still constantly later.
In response to this challenge, crowd workers are instructed to document each behavior from the perspective of triggering conditions ($P1$) and abstract relationships ($P3$).
The documentation of interactive behaviors is not only a specification for end-users (or other designers and developers) to describe these behaviors as part of the sketch, but also a tutorial for the crowd workers that later join the sketch session and should understand/refine/demonstrate the interactive behaviors. 
The documented sketch supports collective memory to help compensate for worker dynamicity in real-time crowd systems \cite{lasecki2011legion,lasecki2012crowd}. 

\vspace{-3pt}
\subsection{Coordination} 
\vspace{-2pt}

Lastly, composing interactive behaviors as a group poses general challenges of real-time groupware, especially given the tasks are open-ended. 
Having the crowd to work collaboratively will have benefits not only in the time it takes to create a interactive behavior as well as making the sketch interactive with multiple complex behaviors that can run concurrently. 
In our example, worker 1 can move the Super Mario character, worker 2 can make Turtle move back and forth, worker 3 observe if Super Mario steps on which turtle to trigger the interactive behavior and so on.
We notice a comparison between having limited tool sets and the ability to lock the activity of demonstrating, remixing and documenting single behavior. 
This locking mechanism prevents conflicts and shares awareness among crowd workers. 
We skip the detailed descriptions of each coordination tool here due to space limitations.

\vspace{-2pt}
\section{Conclusion} 
\vspace{-2pt}
Our demo showcases an extension to Apparition, a crowd-powered prototyping tool for interactive systems. 
Apparition enables an end user to verbally describe interactive behaviors while crowd workers demonstrate new behaviors---or remix prior examples---to create them. 
The extension provides a set of methods and tools to retain the interactive behaviors for further communication and coordination among crowd workers.
Ongoing work aims to learn the structure of interactive behaviors and analyze the crowd's demonstrations and corresponding textual documentation of these behaviors, which can be used to help automate the process.


\bibliography{hcomp_demo_2016}
\bibliographystyle{aaai}

\end{document}